\documentclass[twocolumn,showpacs,aps,amsmath,amssymb,superscriptaddress,prxquantum]{revtex4-2}
\usepackage{amsmath,amssymb,amsfonts,bm}
\usepackage{graphicx,epstopdf}  
\usepackage{color}
\usepackage{xcolor}
\usepackage{extarrows}
\usepackage{enumitem}
\usepackage{empheq}
\usepackage{hyperref}

\usepackage{multirow}
\usepackage{booktabs}

\newcommand{\tr}{{\rm tr}}
\newcommand{\la}{\langle}
\newcommand{\ra}{\rangle}
\newcommand{\trho}{\tilde{\rho}}
\newcommand{\tbrho}{\tilde{\bm \rho}}
\newcommand{\brho}{\bm \rho}
\newcommand{\bR}{\bm{R}}
\newcommand{\btheta}{\bm{\theta}}

\newcommand{\lla}{\la \! \la}
\newcommand{\rra}{\ra \! \ra}
\newcommand{\Pade}{\text{Padé}\ }

\newcommand{\Ht}{H_{_{\rm T}}}
\newcommand{\Hs}{H_{_{\rm S}}}
\newcommand{\He}{H_{_{\rm E}}}
\newcommand{\Hse}{H_{_{\rm SE}}}
\newcommand{\NS}{N_{_{\rm S}}}

\newcommand{\Eq}[1]{Eq.\,(\ref{#1})}

\begin{document}
\title{Error Propagation Theory for Variational Non-Markovian Open Quantum Dynamics}

\author{Long Cao} \email{caolong@mail.ustc.edu.cn}
\affiliation{Hefei National Research Center for Physical Sciences at the Microscale, 
University of Science and Technology of China, Hefei, Anhui 230026, China}

\author{Daochi Zhang}
\affiliation{State Key Laboratory of Porous Materials for Separation and Conversion \& MOE Key Laboratory of Computational Physical Sciences \& Department of Chemistry, Fudan University, Shanghai 200438, China}

\author{Yao Wang}
\affiliation{Hefei National Research Center for Physical Sciences at the Microscale, 
University of Science and Technology of China, Hefei, Anhui 230026, China}

\author{Liwei Ge} 
\affiliation{Hefei National Research Center for Physical Sciences at the Microscale, 
University of Science and Technology of China, Hefei, Anhui 230026, China}

\author{Rui-Xue Xu}
\affiliation{Hefei National Research Center for Physical Sciences at the Microscale, 
University of Science and Technology of China, Hefei, Anhui 230026, China}
\affiliation{Hefei National Laboratory, Hefei, Anhui 230088, China}

\author{YiJing Yan}
\affiliation{Hefei National Research Center for Physical Sciences at the Microscale, 
University of Science and Technology of China, Hefei, Anhui 230026, China}

\author{Xiao Zheng} \email{xzheng@fudan.edu.cn}
\affiliation{State Key Laboratory of Porous Materials for Separation and Conversion \& MOE Key Laboratory of Computational Physical Sciences \& Department of Chemistry, Fudan University, Shanghai 200438, China}
\affiliation{Hefei National Laboratory, Hefei, Anhui 230088, China}

\date{Submitted on August~23, 2026}


\begin{abstract}
    
    Variational approaches based on neural quantum states and physics-informed neural networks provide powerful paradigms for simulating non-Markovian open quantum dynamics. However, extending these methods into the strongly non-Markovian regime reveals a critical bottleneck: even minute errors in the time evolution can translate into substantial deviations in physical observables. The fundamental origin of this stringent precision requirement, as well as how non-Markovianity governs it, remains an open question. 
    Here, we develop a theoretical framework that systematically characterizes error propagation in variational non-Markovian dynamics. By combining analytical derivations with numerical verification, we present the first quantitative description of variational error evolution over time. Our analysis uncovers an intrinsic error-backflow mechanism driven by long-lived environmental memory. This mechanism establishes a fundamental precision barrier and provides concrete guidance for designing robust variational algorithms for strongly non-Markovian quantum systems.
    
\end{abstract}

\maketitle

\section{Introduction}
Many-body open quantum systems (OQSs) have drawn significant attention due to their interdisciplinary relevance, spanning physics, chemistry, materials science, and life sciences. Their applications are remarkably diverse, including coherent energy transfer in biological photosystems \cite{Engel2007,scholes2010quantum,hildner2013quantum}, charge transfer in molecular aggregates \cite{kasha1963energy,hestand2018expanded}, electron transport in single-molecule junctions \cite{10.1063/1.5003306,uzma2021understanding}, multidimensional coherent spectroscopy of condensed-phase materials \cite{10.1063/1.4994987,Bruder2018}, correlated quantum matter for quantum information processing \cite{Monroe2002,head2020quantum}, and high-precision control of local spin states in surface-adsorbed molecules \cite{RevModPhys.94.045008}.

For a realistic simulation of these systems, accurate characterization of environmental dissipations and system correlations is crucial \cite{wilson1975renormalization,madhavan1998tunneling,zhao2005controlling,li2020molecular,zheng2013kondo,doi:10.1126/science.aay6779,doi:10.1126/science.abg8223,Ding25084114}. Consequently, numerous approaches have been developed to simulate the non-Markovian dynamics of many-body OQSs \cite{Tam18030402,Lam193721,strunz1999open,10.1063/1.1647528,han2019stochastic,PhysRevLett.88.170407,suess2014hierarchy,PhysRevLett.115.266802,PhysRevLett.130.186301,PhysRevB.87.115115,PhysRevLett.88.256403,RevModPhys.92.011001,10.1063/1.3173823,RevModPhys.93.015008,li2023dissipatons,Yan14054105,li2024toward,moix2013hybrid,duan2017zero,hsieh2018unified,10.1063/1.1580111,xu2026colloquium}. Among these, the dissipaton-embedded quantum master equation (DQME) \cite{Yan14054105,li2024toward} has emerged as a promising method, offering a compact representation of quantum dissipative dynamics and excellent transferability. Its second-quantized formulation has been demonstrated to be equivalent to the extensively employed hierarchical equations of motion (HEOM) \cite{Tan89101,YAN2004216,Jin08234703,PhysRevLett.109.266403}, and the dissipaton picture shares common features with the pseudomode theory \cite{PhysRevA.55.2290,Tam18030402,Lam193721,cirio2023pseudofermion,Lin251289,PhysRevResearch.2.043058} and bexciton theory \cite{chen2024bexcitonics}.

However, exact numerical methods, including HEOM and DQME, all confront an inherent exponential wall: their computational costs scale exponentially with both the system size and the complexity of environmental memory. To overcome this curse of dimensionality, a common approach is to employ a variational ansatz to compress the time-dependent dynamical variables, such as the reduced density tensor (RDT) in the DQME framework. To this end, tensor network states \cite{verstraete2008matrix,schollwock2011density,orus2014practical,shi2018efficient} and neural quantum states (NQSs) \cite{doi:10.1126/science.aag2302,PhysRevLett.125.100503,PhysRevLett.122.250502,PhysRevLett.122.250501,PhysRevLett.122.250503,PhysRevB.99.214306,cao2026nqs,ye2025simulating} have been developed to efficiently represent these time-dependent quantities. For instance, matrix product states (MPSs) \cite{shi2018efficient,ke2022hierarchical,ke2023tree,preston2025nonadiabatic} have been employed to compress the auxiliary density operators in HEOM, while restricted Boltzmann machines (RBMs) \cite{cao2026nqs} have been used to represent the RDT in DQME. The resulting RBM-DQME and MPS-HEOM approaches have achieved a significant reduction in the number of required dynamical variables. More recently, physics-informed neural networks (PINNs) have emerged as an alternative, offering a potential computational advantage by encoding temporal evolution directly into the network inputs and optimizing residuals from the governing equations, thereby bypassing the costly time-step updates required by the traditional time-dependent variational principle (TDVP) in NQSs \cite{cao2026pinn}. 
Parallel to these conventional variational frameworks, operator learning techniques \cite{zhang2024artificial,zhang2025neural} and quantum neural networks \cite{long2024quantum} have also emerged as promising paradigms for simulating the dissipative dynamics of OQSs.

Both NQS-DQME and PINN-DQME methods exhibit remarkable precision and efficiency in the high-temperature, weakly non-Markovian regime \cite{cao2026nqs,cao2026pinn}. Yet, both approaches encounter a severe shared bottleneck when extended to the strongly non-Markovian regime. As the environmental temperature decreases, the bath correlation time grows substantially, pushing the open quantum dynamics deep into this challenging domain. Crucially, it was demonstrated that physical observables extracted from the RDT in NQS-DQME can exhibit appreciable errors even when TDVP yields negligible residual errors \cite{cao2026nqs}. This reveals a steep reduction in the error tolerance of the underlying numerical procedures, specifically TDVP projections for NQSs and loss optimizations for PINNs. Consequently, preserving observable accuracy demands exceptionally tight precision thresholds for the time evolution of the RDT \cite{cao2026nqs,cao2026pinn}. This implies that in long-memory environments, even near-exact variational projections or loss minimization cannot guarantee the accuracy of predicted physical observables.

While existing theoretical efforts have overwhelmingly focused on establishing the static representational capacity and expressivity of variational ansatzes \cite{deng2017quantum,carleo2019machine,sharir2022neural,levine2019quantum}, a fundamental understanding of how errors accumulate and propagate during time-dependent
variational evolution remains nascent \cite{schmitt2025simulating}. In this work, we bridge this critical gap by developing a comprehensive theoretical framework for variational error propagation in non-Markovian open quantum dynamics. We unveil the core physical mechanism driving this error accumulation: non-Markovianity acts as a memory channel that gives rise to an ``error backflow'', wherein numerical errors incurred at earlier evolution times are reinjected into the system via the bath, leading to a drastic amplification of observable errors over the course of evolution.


By combining analytical derivations with concrete numerical validations, our theory quantitatively characterizes temporal error bounds and pinpoints the physical origin of the severe precision requirements in strongly non-Markovian regimes. Ultimately, this work clarifies the fundamental failure mechanisms of current variational dynamics and provides clear theoretical guidelines for designing error-resilient algorithms in strong-memory quantum environments.

The remainder of this paper is organized as follows. In Section~\ref{subsec:dqme}, we provide the details of the DQME method and TDVP. In Section~\ref{subsec:error}, we present
the theoretical analysis of the variational error accumulation.
Section~\ref{sec:num} presents numerical results and discusses the underlying physical mechanisms. Section~\ref{sec:conclude} concludes with a summary and outlook.

\section{Methodology} \label{sec:method}
\subsection{Fermionic DQME formalism and time-dependent variational principle}  \label{subsec:dqme}

We consider an Anderson impurity model \cite{And6141} in which the impurity is linearly coupled to the environment consisting of several noninteracting electron reservoirs. 
The Hamiltonian of the total system is described by (with $e = \hbar  = 1$)
\begin{align}
\Ht & = \Hs + \He + \Hse, \\
\He & = \sum_\alpha \sum_{k,s}  \epsilon_{\alpha ks} \hat{d}^{\dagger}_{\alpha ks}\hat{d}_{\alpha ks}, \\
\Hse & = \sum_{\nu=1}^{N_{_{\nu}}} \sum_{\alpha, s}
	\hat{c}_{\nu s}^{\dagger}\hat{F}_{\alpha \nu s} + \hat{F}_{\alpha \nu s}^{\dagger}\hat{c}_{\nu s}, 
\end{align}
where $\Hs$ represents the system of primary interest, $\He$ denotes the environment, and $\Hse$ describes the system-environment coupling. 
In these terms, $\hat{c}^\dag_{\nu s}$ ($\hat{c}_{\nu s}$) denotes the creation (annihilation) operator for an electron of spin $s$ at the $\nu$-th energy level of the system,  $\hat{d}^\dag_{\alpha k s}$ ($\hat{d}_{\alpha k s}$) denotes the creation (annihilation) operator for an electron of spin $s$ at the $k$-th state of the $\alpha$-th reservoir. 
%
$\hat{F}_{\alpha \nu s}=\sum_k t_{\nu \alpha k s}\hat{d}_{\alpha k s}$, where $t_{\nu \alpha k s}$ denotes the coupling strength between the system's $\nu$-th state and the $\alpha$-reservoir's $k$-th state for the electrons with spin $s$.
%

Since a reservoir of  noninteracting fermionic particles linearly coupled to the system follows Gaussian statistics, its influence on the reduced system dynamics is entirely captured by the reservoir hybridization correlation functions \cite{Jin08234703}, which are defined as
\begin{align}
    C_{\alpha \nu s}^{+}(t-\tau) & \equiv \langle 
        \hat{F}_{\alpha \nu s}^{\dagger}(t) \hat{F}_{\alpha \nu s}(\tau) \rangle_{_{\rm E}}, \nonumber \\ 
    C_{\alpha \nu s}^{-}(t-\tau) & \equiv \langle
        \hat{F}_{\alpha \nu s}(t) \hat{F}^{\dagger}_{\alpha \nu s}(\tau) \rangle_{_{\rm E}},
\end{align}
where $\hat{F}_{\alpha \nu s}(t) = e^{i \He t}\hat{F}_{\alpha \nu s}e^{-i \He t}$, and $\la \cdot \ra_{_{\rm E}} = {\rm tr}( \cdot \rho^{\rm eq}_{_{\rm E}})$ with $\rho^{\rm eq}_{_{\rm E}}$ being the density matrix of the decoupled environment in thermal equilibrium. 
The reservoir correlation functions are related to the hybridization spectral functions through the fluctuation-dissipation theorem
\begin{equation}
    C_{\alpha \nu s}^{\sigma}(t) = \frac{1}{\pi}
    \int_{-\infty}^{\infty} {\rm d} \omega \, e^{\sigma i \omega t} 
    f^{\sigma}_{\alpha}(\omega) J_{\alpha \nu s}(\omega), \label{eqn:FD}
\end{equation}
where $f^{\sigma}_{\alpha}(\omega) = 1/(1+e^{\sigma \beta_{\alpha} (\omega-\mu_{\alpha})})$, and $\beta_{\alpha} = 1/(k_{\rm B}T)$ is the inverse temperature of the $\alpha$-reservoir. 
The hybridization spectral density functions of the $\alpha$-reservoir are defined as
\begin{equation}
    J_{\alpha \nu s}(\omega) \equiv \pi \sum_k \left\lvert t_{\nu \alpha k s} \right\rvert^2
    \delta(\omega-\epsilon_{\alpha k s}).
\end{equation}
In this work, we adopt a Lorentzian form for the hybridization spectral functions, i.e.,
\begin{equation}
    J_{\alpha \nu s}(\omega) = \frac{\Gamma_{\nu s\alpha} W_\alpha^2 }{(\omega-\Omega_{\alpha})^2+W_\alpha^2}.  
\end{equation}
Here, $\Gamma_{\nu s \alpha}$ is the hybridization strength between the system's $\nu$-th state and the $\alpha$-reservoir for the electrons with spin $s$, $\Gamma_{\nu s} = \sum_\alpha \Gamma_{\nu s \alpha}$, and $\Omega_\alpha$ and $W_\alpha$ are the band center and width of the $\alpha$-reservoir, respectively. 
For simplicity, we take $\Omega_\alpha = \mu_\alpha$, where $\mu_\alpha$ is the chemical potential of the $\alpha$-reservoir. In particular, we set $\mu_\alpha^{\rm eq} = 0$ in thermal equilibrium.

By employing a sum-over-poles expansion for $f^{\sigma}_{\alpha}(\omega)$ and $J_{\alpha \nu s}(\omega)$ in Eq.~\eqref{eqn:FD}, the reservoir correlation functions can be expressed as a linear combination of exponential functions
\begin{equation}
	C_{\alpha \nu s}^{\sigma}(t) = \sum_{p=0}^{N_{_{\rm M}}-1} 
    \eta_{\alpha \nu s p}^{\sigma} e^{-\gamma_{\alpha \nu sp}^{\sigma}t},
    \label{eqn:Ct-exp-1}
\end{equation}
where $N_{_{\rm M}}$ is the total number of decomposition modes, and the index $p$ labels the distinct modes.
To be concise, we define the $j$ index as the multi-index $\{\alpha \nu s p\}$ here. We adopt the \Pade spectral decomposition scheme \cite{hu2010communication} to perform the exponential expansion in \Eq{eqn:Ct-exp-1}. In the \Pade spectral decomposition scheme, we have $N_{_{\rm M}}=N_{\Pade}+1$, where $N_{\Pade}$ is the number of \Pade poles associated with the reservoir spectral functions.

\begin{figure}[t]
\includegraphics[width=\columnwidth]{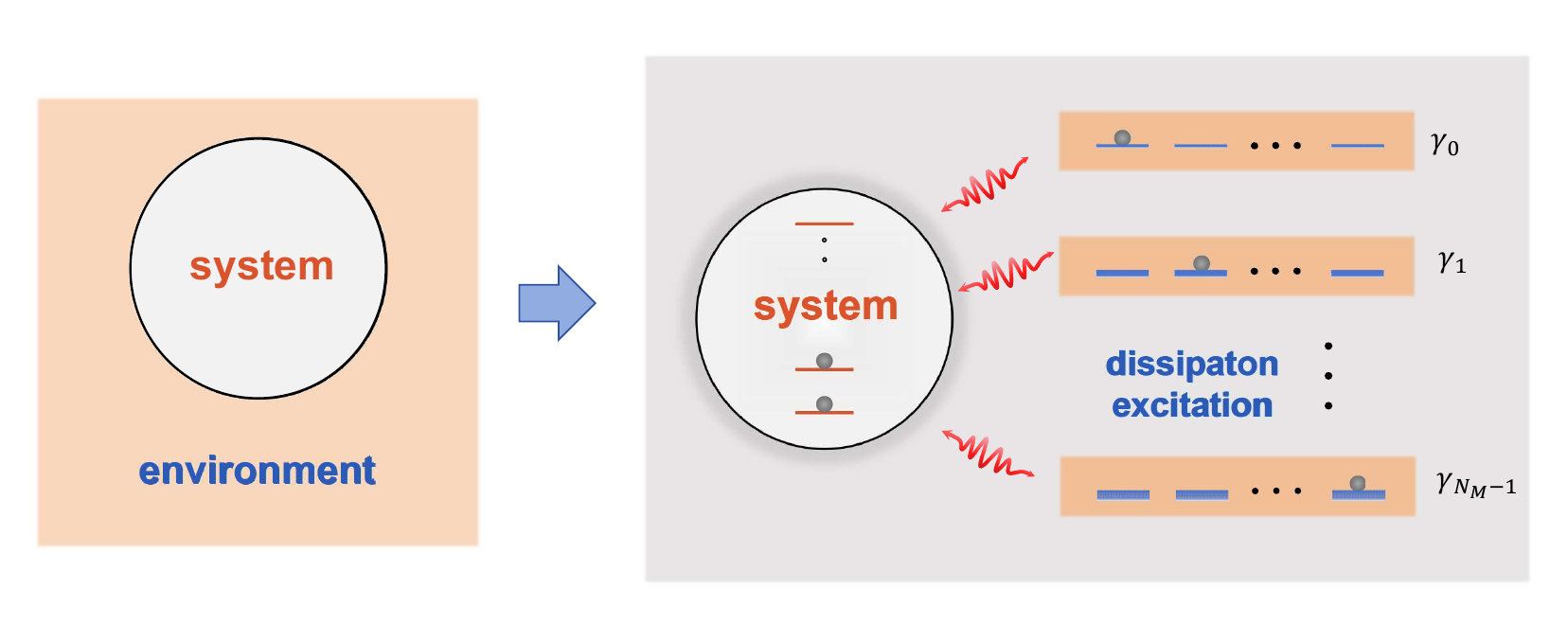}
	\caption{Schematic of the fermionic DQME theory, mapping the original OQS (left) to a dissipaton-embedded system (right). Red and blue bars represent the $\NS$ system fermion energy levels and $N_{_{\rm E}}$ memory-carrying dissipaton levels, respectively. 
    Broadening of blue bars indicates each dissipaton's decay rate (inverse lifetime). The dissipaton levels can be grouped into $N_{_{\rm M}}$ decay channels, where the levels within each channel share a common decay rate. } \label{fig1}
\end{figure}

In the second quantized formulation of DQME, the expansion of the reservoir correlation functions defines the dissipatons, which are dissipative quasiparticles with their complex energies $\gamma_j^\sigma$ describing the non-Markovian effects \cite{hu2010communication,Yan14054105,li2024toward}. Here,
$j$ labels the dissipaton levels, and $\sigma$ denotes the dissipaton charge ($\sigma=+$ for hole-type; $\sigma=-$ for electron-type). The imaginary and real parts of $\gamma_j^\sigma$ correspond to the dissipaton energy and inverse lifetime (or decay rate), respectively, while $\eta^\sigma_j$ quantifies the system-dissipaton coupling strength. 
In many cases, the decay rate depends solely on the decomposition mode $p$. Consequently, these decomposition modes are also referred to as the decay channels of dissipatons.

Figure~\ref{fig1} illustrates the fermionic DQME theory, where the original OQS is mapped to a dissipaton-embedded system characterized by the reduced density tensor (RDT), $\tbrho = \{\trho(\vec{n},\vec{n}';\vec{m}^-,\vec{m}^+)\}$. Here, $\trho(\vec{n},\vec{n}';\vec{m}^-,\vec{m}^+)=\lla\vec{n},\vec{n}';\vec{m}^-,\vec{m}^+\vert\tbrho\rra$, $\vert \vec{n},\vec{n}';\vec{m}^-,\vec{m}^+ \rra=\vert \vec{n},\vec{n}'\rra_{_{\rm S}} \otimes \vert\vec{m}^-,\vec{m}^+ \rra_{_{\rm E}}$ are the basis vectors of the extended Liouville space composed of the system Liouville space (spanned by $\{\vert \vec{n},\vec{n}'\rra_{_{\rm S}}\}$) and the dissipaton space (spanned by $\{\vert\vec{m}^-,\vec{m}^+ \rra_{_{\rm E}}\}$). Vectors $\vec{n}$ and $\vec{n}'$ denote system fermion configurations, while $\vec{m}^-$ and $\vec{m}^+$ represent electron-type and hole-type dissipaton configurations, respectively. The RDT elements with $\vert\vec{m}^-,\vec{m}^+\rra_{_{\rm E}} = \vert\vec{0}\rra_{_{\rm E}}$ (the dissipaton vacuum) yield the system's reduced density operator (RDO): $\brho_{_{\rm S}} = \lla\vec{0}\vert_{_{\rm E}} \tbrho \rra$. 
For convenience, we simplify the notation by omitting the subscript of the left vector, writing $\lla \cdot \vert_{_{\rm E}} \cdot \rra_{_{\rm E}}$ as $\lla \cdot \vert \cdot \rra_{_{\rm E}}$ in the following sections.
As shown in Fig.~\ref{fig1}, each level can accommodate at most one fermion or dissipaton in accordance with the Pauli exclusion principle. 
In practice, a maximal dissipaton occupation number $M_{\rm max}$ is introduce to facilitate the numerical solution of the DQME \cite{li2024toward}.

The fermionic DQME is given by \cite{li2024toward}
\begin{align}
        \dot{\tbrho} =\mathcal{L}\tbrho=& -i[H_{_{\rm S}},\tbrho] - \sum_j \left(
        \gamma_j^- \hat{N}_j \tbrho + \gamma_j^+ \tbrho \hat{N}_j\right) \notag \\
       & -i \sum_j \Big[ \big(\hat{c}_{\nu}^{\dagger} \hat{b}_{j} \tbrho - \hat{b}_{j} \tbrho \hat{c}_{\nu}^{\dagger} \big) + 
       \big(\hat{c}_{\nu} \tbrho \hat{b}^{\dagger}_j -  \tbrho \hat{b}_{j}^{\dagger} \hat{c}_{\nu} \big) \Big] \notag \\
       & -i \sum_j  \left[-\eta_{j}^{-}\, \hat{c}_{\nu} \hat{b}_j^{\dagger} \, \tbrho - (\eta_{j}^{+})^\ast \,\hat{b}_j^{\dagger}\tbrho \hat{c}_{\nu}\right]  \notag \\
       & -i \sum_j \left[ 
       \eta_{j}^{+} \, \hat{c}_{\nu}^{\dagger}\tbrho\, \hat{b}_j
       + (\eta_{j}^{-})^\ast  \tbrho\, \hat{b}_j \hat{c}_{\nu}^{\dagger}
       \right]. \label{eqn:DQME-SQ-1}
\end{align}
Here, $\hat{c}_{\nu}^{\dagger}$ ($\hat{c}_{\nu}$) creates (annihilates) a fermion at the $\nu$-th system fermion level, $\hat{b}_j^{\dagger}$ ($\hat{b}_j$) creates (annihilates) a dissipaton at $j$-th level, and $\hat{N}_j = \hat{b}_j^{\dagger} \hat{b}_j$ gives the dissipaton occupation. 


Fundamentally, dissipatons share similarities with previously proposed pseudomodes \cite{PhysRevA.55.2290,Tam18030402,Lam193721,cirio2023pseudofermion,Lin251289,PhysRevResearch.2.043058}, and the DQME is formally equivalent to the exact HEOM theory \cite{Jin08234703,PhysRevLett.109.266403} for reduced system dynamics.


We consider the DQME as $\dot{\tbrho}=\mathcal{L}\tbrho$, 
and represent the RDT with a variational ansatz $\tbrho_{\btheta}$, where $\btheta$ denotes the variational parameters. At time $t$, TDVP projects the tangent vector $\mathcal{L}\tbrho_{\btheta}(t)$ of the extended Liouville space that represents the time increment in $\tbrho_{\btheta}(t)$ to the parameter tangent space to get the evolution equation of parameters
\begin{align}
    \dot{\btheta} = \arg\min_{\dot{\btheta}} \Vert
   \dot{\tbrho}_{\btheta}-\mathcal{L}\tbrho_{\btheta}(t)
    \Vert,
\end{align}
where $\dot{\tbrho}_{\btheta}=\sum_k\dot{\theta}_k \partial_k\tbrho_{\btheta}$, and $\Vert \cdot \Vert$ denotes the norm defined on the extended Liouville space. In the following, we adopt the 2-norm of the vector in Liouville space, defined as $\Vert\tbrho\Vert=\Vert\tbrho\Vert_2=\sqrt{\sum_j \vert\tilde{\rho}_j\vert^2}$.

\subsection{Error propagation within the NQS-DQME framework}
\label{subsec:error}
Throughout the following derivation, we consider a system where the Liouvillian superoperator $\mathcal{L}$ is time-independent, while the main conclusions can be easily generalized to time-dependent systems (see Appendix~\ref{app:tl} for details).

Firstly, we define
\begin{align}
\delta \tbrho_{\btheta}(t) & = \tbrho_{\btheta}(t)-\tbrho(t) ,
\end{align}
where $\tbrho(t)$ is the exact RDT at time $t$, $\tbrho_{\btheta}(t)$ is the parameterized representation of $\tbrho(t)$, whose dynamics is governed by the TDVP equation. Taking the time derivative of this equation, we get
\begin{align}
\delta \dot{\tbrho}_{\btheta}(t) & =\dot{\tbrho}_{\btheta}(t)-
\mathcal{L}\tbrho_{\btheta}(t)+\mathcal{L}\tbrho_{\btheta}(t)-\mathcal{L}\tbrho(t)
\notag \\
&= \bR(t) +\mathcal{L} \delta \tbrho_{\btheta}(t).
\end{align}
Here, $\bR(t)=\dot{\tbrho}_{\btheta}(t)-\mathcal{L}\tbrho_{\btheta}(t)$ denotes the error vector of the TDVP projection, comprising the approximation error, regularization error, and sampling error. Analytically solving this equation, we have (by setting $t_0 = 0$)
\begin{align}
\delta \tbrho_{\btheta}(t) &= e^{\mathcal{L}t} \delta \tbrho_{\btheta}(0) + \int_{0}^{t} dz \, e^{\mathcal{L}(t-z)} \bR(z).
\label{eq:deltarho}
\end{align}
Whether errors primarily stem from initial state deviations $\delta \tbrho_{\btheta}(0)$ or dynamical variational errors $\bR(z)$ depends on the choice of the initial time. Focusing here on the dynamic error accumulation during the TDVP evolution, we assume zero initial error, $\delta \tbrho_{\btheta}(0)=0$, and presume that the projection error is constant over time, i.e., $\bR(t)\approx\bR$.
This yields:
\begin{equation}
\delta \tbrho_{\btheta}(t) \approx  \int_{0}^{t} dz \, e^{\mathcal{L}(t-z)} \bR. \label{deltarho}
\end{equation}
For any system observable quantity $\hat{O}$, we thus have
\begin{align}
    \langle \delta\hat{O} (t)\rangle_{\btheta} &=
     \tr_{_{\rm S}} \big[\hat{O} \,
     \lla \vec{0} \vert \delta\tbrho_{\btheta}(t)
      \rra_{_{\rm E}} \big]
     \notag \\
     &\approx \tr_{_{\rm S}} \Big[\hat{O} \,
      \lla \vec{0} \vert \int_{0}^{t} dz \, e^{\mathcal{L}(t-z)} \bR\rra_{_{\rm E}}\Big],
\end{align}
where $\tr_{_{\rm S}}$ denotes the partial trace over the system degrees of freedom, 
Inserting the completeness of dissipaton levels $\sum_{\vec{m}}\vert\vec{m}\rra_{_{\rm E}}\lla\vec{m}\vert_{_{\rm E}}=\bm{1}_{_{\rm E}}$, 
where $\vert \vec{m} \rra_{_{\rm E}}=\vert\vec{m}^-,\vec{m}^+\rra_{_{\rm E}}$, 
we have
\begin{align}
    \langle \delta\hat{O}(t)\rangle_{\btheta} &=\sum_{\vec{m}}
      \tr_{_{\rm S}} \Big[\hat{O} \,
      \lla \vec{0} \vert \int_{0}^{t} dz \, e^{\mathcal{L}(t-z)}\vert\vec{m}\rra_{_{\rm E}} \lla\vec{m}\vert \bR
      \rra_{_{\rm E}} \Big] \nonumber \\
      & \quad \leqslant\sum_{\vec{m}}\Vert\hat{O}\Vert_{F}\cdot
     \left\Vert\lla\vec{0}\vert\int_{0}^{t} dz
     \, e^{\mathcal{L}(t-z)}\vert\vec{m}
    \rra_{_{\rm E}}\right\Vert_{F}  \nonumber \\
    & \qquad  \qquad  \cdot    \left\Vert\bR_{\vec{m}}\right\Vert_{F}, \label{eq:deltaQ}
\end{align}
where $\Vert A\Vert_{F}  = \sqrt{\operatorname{Tr}(A^{\dagger} A)}$ denotes the Frobenius norm of the matrix $A$, and $\bR_{\vec{m}}=\lla\vec{m}\vert \bR\rra_{_{\rm E}}$. Here, $\hat{O}$ and $\bR_{\vec{m}}$ are $\NS\times \NS$ matrices, while $\lla\vec{0}\vert\int_{0}^{t} dz
     \, e^{\mathcal{L}(t-z)}\vert\vec{m}
    \rra_{_{\rm E}}$ is a $\NS^2\times \NS^2$ matrix.

The inequality in Eq.~\eqref{eq:deltaQ} is deduced by applying the Cauchy-Schwarz inequality twice. First, applying the Frobenius-norm Cauchy-Schwarz inequality $\vert \tr [AB]\vert\leqslant \Vert A\Vert_F \Vert B\Vert_F $ to the $\NS\times \NS$ matrices $A=\hat{O}
      \lla \vec{0} \vert \int_{0}^{t} dz \, e^{\mathcal{L}(t-z)}\vert\vec{m}\rra_{_{\rm E}}$ and $B=\lla\vec{m}\vert \bR  \rra_{_{\rm E}}$, we obtain
\begin{align}
    & \tr_{_{\rm S}} \Big[\hat{O} \,
      \lla \vec{0} \vert \int_{0}^{t} dz \, e^{\mathcal{L}(t-z)}\vert\vec{m}\rra_{_{\rm E}}
      \lla\vec{m}\vert \bR  \rra_{_{\rm E}} \Big]
      \notag \\
      \qquad &\leqslant
      \left\Vert \hat{O}
      \lla \vec{0} \vert \int_{0}^{t} dz \, e^{\mathcal{L}(t-z)}\vert\vec{m}\rra_{_{\rm E}}\right\Vert_F \cdot
      \left\Vert \lla\vec{m}\vert \bR 
      \rra_{_{\rm E}}\right\Vert_F. 
\end{align}
The equality condition here is $\lla\vec{m}\vert \bR\rra_{_{\rm E}}=\lambda [\hat{O}
      \lla \vec{0} \vert \int_{0}^{t} dz \, e^{\mathcal{L}(t-z)}\vert\vec{m}\rra_{_{\rm E}}]^{\dagger}=\lambda \lla \vec{m} \vert \int_{0}^{t} dz \, e^{\mathcal{L^\dagger}(t-z)}\vert\vec{0}\rra_{_{\rm E}}\hat{O}^{\dagger}$, where $\lambda$ is a complex number.
As $\lla\vec{m}\vert \bR\rra_{_{\rm E}}$ is random, this condition is attainable, implying that our estimation is not overly conservative.

Second, vectorizing the matrix $\hat{O}$ as a $\NS^2$ vector $x={\rm vec}(\hat{O})$ and applying $\Vert Ax\Vert_2 \leqslant \Vert A\Vert_F \Vert x\Vert_2$ to the $\NS^2\times \NS^2$ matrix $A=\lla \vec{0} \vert \int_{0}^{t} dz \, e^{\mathcal{L}(t-z)}\vert\vec{m}\rra_{_{\rm E}}$, we have
\begin{align}
     &\left\Vert \hat{O}
      \lla \vec{0} \vert \int_{0}^{t} dz \, e^{\mathcal{L}(t-z)}\vert\vec{m}\rra_{_{\rm E}}\right\Vert_F 
      \notag \\
      &=
           \left\Vert {\rm vec}\left(\hat{O}
      \lla \vec{0} \vert \int_{0}^{t} dz \, e^{\mathcal{L}(t-z)}\vert\vec{m}\rra_{_{\rm E}}\right)\right\Vert_2
      \notag \\
      &\leqslant
      \left\Vert {\rm vec}(\hat{O}) \right\Vert_2 \cdot
      \left\Vert\lla \vec{0} \vert \int_{0}^{t} dz \, e^{\mathcal{L}(t-z)}\vert\vec{m}\rra_{_{\rm E}}\right\Vert_F,
\end{align}
where $\Vert x \Vert_2=\sqrt{\sum_ix_i^2}$ denotes the 2-norm of the vector $x$, and we have
$\Vert \hat{O} \Vert_F = \Vert {\rm vec}(\hat{O}) \Vert_2$. Equality in the inequality (i.e., saturation of the bound) requires two conditions: (i) $A = \lla \vec{0} \vert \int_{0}^{t} dz \, e^{\mathcal{L}(t-z)}\vert\vec{m}\rra_{_{\rm E}}$ is a rank-1 matrix, and (ii)  
$x = {\rm vec}(\hat{O}) \propto v_1^\dagger$, where $v_1$ is 
the non-zero left singular vector of $A$.
The first condition is approximately attainable because, as $t\to +\infty$, $I(t)=\int_{0}^{t} dz \, e^{\mathcal{L}(t-z)}$ is dominated by the smallest non-zero eigenvalue of $-\mathcal{L}$,
\begin{align}
    I(t) &= \sum_{k=0} \left( \int_{0}^{t} e^{\lambda_k \tau} d\tau \right) \vert \psi_k \rra \lla \phi_k \vert 
    \notag \\
    &= t\vert \psi_0 \rra \lla \phi_0 \vert + \sum_{k=1} \frac{e^{\lambda_k t} - 1}{\lambda_k} \vert \psi_k \rra \lla \phi_k \vert,
\end{align}
where $\vert \psi_k \rra$ and $\vert \phi_k \rra$ are the right and left eigenvectors of $\mathcal{L}$, respectively, with $k=0$ corresponding to the steady state. When $t \to +\infty$, 
\begin{enumerate}
    \item Although the steady-state mode ($\lambda_0 = 0$) accumulates a weight that grows linearly with time $t$, the overlap $\lla \phi_0 \vert\vec{m}\rra_{\rm E}$ is negligibly small for any non-zero index $\vec{m} \neq \vec{0}$. Meanwhile, the $\vec{m} = \vec{0}$ component merely contributes a trivial inter-system correlation, which is inconsequential to our error analysis (see Appendix~\ref{app:reg} for details).
    

    \item The other states have the weight $1/\vert\lambda_k\vert$, and are dominated by the first term $1/\vert\lambda_1\vert$ which is smallest among all $1/\vert\lambda_k\vert$ with $k\geqslant 1$.
    
\end{enumerate}

The second condition is also attainable because the observable $\hat{O}$ is not confined to a specific form.

Suppose that $\Vert\bR_{\vec{m}}\Vert_{F}$ has a maximum $r_{\rm max}$, we have the upper error bound of the system observable quantity
\begin{align}
    \frac{\langle \delta\hat{O}(t)\rangle_{\btheta}}{\Vert\hat{O}\Vert_{F}}
    &\leqslant
     r_{\rm max}\cdot
     \sum_{\vec{m}}\left\Vert\lla\vec{0}\vert\int_{0}^{t} dz \, e^{\mathcal{L}(t-z)} \vert \vec{m}\rra_{_{\rm E}}
    \right\Vert_{F}
    \notag \\
    &=r_{\rm max}\cdot
     \chi_0(t)\cdot\chi(t).\label{errordqme}
\end{align}
Here, $\chi(t)$ is the non-Markovian error susceptibility, which links the instantaneous TDVP error to the cumulative error in observable quantities. It is defined as
\begin{align}
\chi(t) &= \sum_{m=0}^{M_{\rm max}}\chi_m(t)/\chi_0(t),\\
    \chi_m(t) &= \sum_{\sum_j m_j =m}
\left\Vert\lla\vec{0}\vert\int_{0}^{t} dz \, e^{\mathcal{L}(t-z)} \vert \vec{m}\rra_{_{\rm E}}
    \right\Vert_{F}.
\end{align}
Essentially, $\chi(t)$ acts as an amplification factor on top of the Markovian error susceptibility $\chi_0(t)$.


We can see from Eq.~\eqref{eq:deltaQ} that the quantity $\lla\vec{0}\vert\int_{0}^{t} dz \, e^{\mathcal{L}(t-z)} \vert \vec{m}\rra_{_{\rm E}}$ describes how the error flows from the state $\vert \vec{m}\rra_{_{\rm E}}$ to the state $\vert \vec{0}\rra_{_{\rm E}}$. 
Consequently, $\chi_0 (t)$ estimates the error accumulated within the system subspace, whereas $\chi_m (t)\,(m>0)$ quantifies the error propagated from dissipaton configurations with a total occupation number of $i$. $\chi (t)$ denotes the ratio between the total error and the intra-system error, acting as a dynamic noise amplifier: when $\chi(t) \gg 1$, even an extremely small optimization residual $r_{\rm max}$ in the variational ansatz can lead to huge errors in physical observables.This relation demonstrates that as the susceptibility $\chi(t)$ increases, the maximum allowable residual $r_{\rm max}$ for a given precision target in $\hat{O}$ decreases, leading to significantly stricter precision requirements.

On the other hand, $\lla\vec{0}\vert\int_{0}^{t} dz \, e^{\mathcal{L}(t-z)} \vert \vec{m}\rra_{_{\rm E}}$ with $\vec{m}\neq \vec{0}$ inherently stems from non-Markovian effects. Therefore, $\chi(t)$ can also serve as a measure of the system’s non‑Markovianity. 
This becomes evident when comparing Eq.~\eqref{errordqme} with the corresponding results from the Lindblad equation or any other quantum master equation governing solely the time evolution of the RDO:
\begin{equation}
    \frac{\langle \delta\hat{O}(t)\rangle_{\btheta}}{\Vert\hat{O}\Vert_{F}}
    \leqslant
     r_{\rm max}\cdot
     \left\Vert\int_{0}^{t} dz \, e^{\mathcal{L}(t-z)}\right \Vert_{F}=r_{\rm max}\cdot
     \chi_0(t).
\end{equation}

Consequently, the non-Markovian susceptibility $\chi(t)$ bridges the system's intrinsic non-Markovianity and the precision threshold required for accurate observable evaluations, dictating that stronger memory effects inevitably demand higher computational precision. This relationship will be systematically verified via numerical calculations in the subsequent sections. Crucially, the explicit mathematical structure of $\chi(t)$ unveils the underlying physical mechanism of this interplay: within non-Markovian dynamics, the backflow from the environment to the system carries not only physical information but also accumulated numerical errors.




\section{Numerical Results}
\label{sec:num}

In this section, we numerically investigate the dependence of $\chi_m(t)$ on temperature, environmental energy structure, and system degrees of freedom. Furthermore, we pinpoint the dominant environmental channels that dictate the error backflow process.


\subsection{Temperature dependence of $\chi(t)$ }

\begin{figure}[!ht]
	\includegraphics[width=\columnwidth]{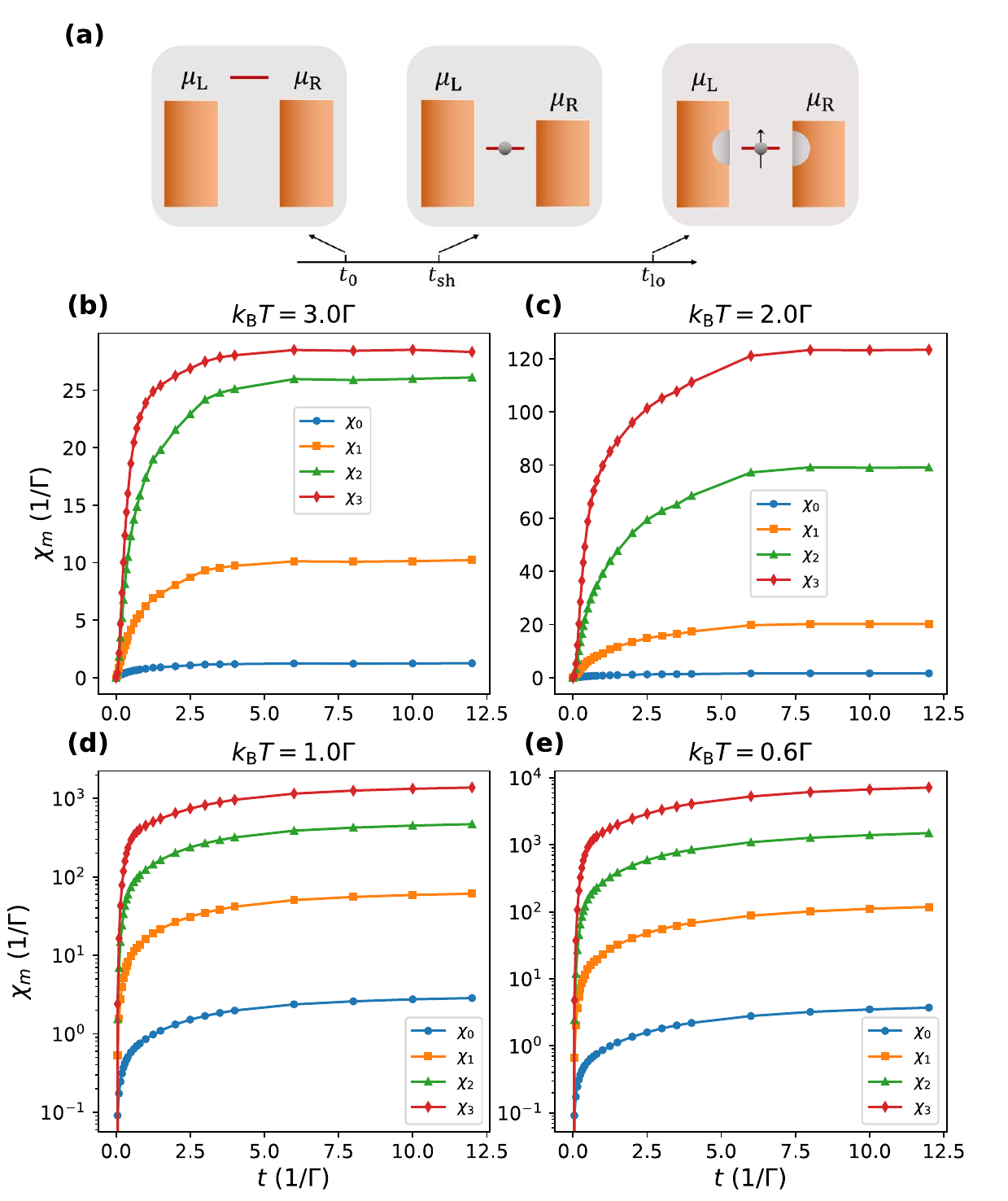}
	\caption{
(a) Schematic of an open quantum system consisting of an impurity coupled to left ($L$) and right ($R$) reservoirs with chemical potentials $\mu_L$ and $\mu_R$. The gray shaded regions in the reservoirs denote Kondo clouds that screen the impurity's localized spin.  
 (b) The trajectories of $\chi_m(t)\ (m=0,1,2,3)$ at $k_{\rm B}T = 3.0\,\Gamma$, with $M_{\rm max} = 3$, $N_{_{\rm M}}=4$, and $\operatorname{Re}(\gamma)_{\rm min}=9.42\,\Gamma$. 
  (c) The trajectories of $\chi_m(t)\ (m=0,1,2,3)$ at $k_{\rm B}T = 2.0\Gamma$, with $M_{\rm max} = 3$, $N_{_{\rm M}}=5$, and $\operatorname{Re}(\gamma)_{\rm min}=6.28\,\Gamma$. 
   (d) The trajectories of $\chi_m(t)\ (m=0,1,2,3)$ at $k_{\rm B}T = 1.0\Gamma$, with $M_{\rm max} = 3$, $N_{_{\rm M}}=6$, and $\operatorname{Re}(\gamma)_{\rm min}=3.14\,\Gamma$. 
    (e) The trajectories of $\chi_m(t)\ (m=0,1,2,3)$ at $k_{\rm B}T = 0.6\Gamma$, with $M_{\rm max} = 3$, $N_{_{\rm M}}=7$, and $\operatorname{Re}(\gamma)_{\rm min}=1.88\,\Gamma$. 
 The vertical axes in (d) and (e) are on a logarithmic scale.
    System parameters (in units of $\Gamma$): $\epsilon_0 = U_0/2 = 2$, $\Delta\epsilon = -7$, and $\Delta U = 6$.
    }  \label{fig:chivsT}
\end{figure}

\begin{table}[ht]
\centering
\caption{The value of $\chi_m(t)$ at $t_{\rm asy}=20.\,\Gamma^{-1}$ for different temperatures.}  \label{tab:chivsT}
\begin{tabular*}{\columnwidth}{@{\extracolsep\fill}lccccc}
\hline\hline
$k_{\rm B}T/\Gamma$  & $3.0$  & $2.0$  & $1.0$  & $0.6$ \\
\hline
$\chi_0(t_{\rm asy})/(1/\Gamma)$  & $1.24$  & $1.66$  & $2.91$  & $4.06$ \\ 
$\chi_1(t_{\rm asy})/(1/\Gamma)$  & $1.01\times 10^{1}$  & $2.02\times 10^{1}$  & $6.26\times 10^{1}$  & $1.29\times 10^{2}$ \\ 
$\chi_2(t_{\rm asy})/(1/\Gamma)$  & $2.60\times 10^{1}$  & $7.90\times 10^{1}$  & $4.82\times 10^{2}$  & $1.63\times 10^{3}$ \\ 
$\chi_3(t_{\rm asy})/(1/\Gamma)$   & $2.85\times 10^{1}$  & $1.23\times 10^{2}$  & $1.41\times 10^{3}$  & $7.81\times 10^{3}$  \\
\hline\hline
\end{tabular*}
\end{table}

\begin{table}[ht]
\centering
\caption{Number of \Pade poles and environmental degrees of freedom in Figure~\ref{fig:chivsT}.}  \label{tab:poleTs}
\begin{tabular*}{\columnwidth}{@{\extracolsep\fill}lccccc}
\hline\hline
$k_{\rm B}T/\Gamma$  & $3.0$  & $2.0$  & $1.0$  & $0.6$  \\
\hline
$N_{\Pade}$  & 3  & 4  & 5  & 6  \\
$N_{\rm M}$\footnote{The values of $N_{\rm M}$ chosen here ensure fully convergent numerical results for all physical observable quantities in this case.}  & 4  & 5  & 6 & 7   \\
$N_{\rm E}$  & 32 & 40 & 48 & 56  \\
$\operatorname{Re}(\gamma)_{\rm min}/\Gamma$ & $9.42$  & $6.28$  & $3.14 $  & $1.88$ \\
\hline\hline
\end{tabular*}
\end{table}

We consider an OQS comprising a localized impurity symmetrically coupled to two noninteracting electron reservoirs. This setup is important for understanding electron transport through molecular junctions. The OQS is described by the single-impurity Anderson model \cite{And6141}. The impurity Hamiltonian is
$
H_{_{\rm S}}(t) = \epsilon_{0} (\hat{n}_\uparrow + \hat{n}_\downarrow)  + U_0\hat{n}_{\uparrow}\hat{n}_{\downarrow} + 
{\Theta}(t-t_0) [\Delta \epsilon (\hat{n}_\uparrow + \hat{n}_\downarrow) + \Delta U\hat{n}_{\uparrow}\hat{n}_{\downarrow}]
$,
where $\hat{n}_s$ is the occupation operator for spin-$s$ electrons, $\epsilon_0$ is the impurity energy, $U_0$ is the Coulomb interaction energy, and $\Theta(t)$ denotes a step function. We consider the following non-Markovian process:
at time $t_0 = 0$, a sudden quench shifts the impurity's parameters by $\Delta \epsilon$ and $\Delta U$, and a bias voltage is applied, establishing a chemical potential difference between the reservoirs. 
Figure~\ref{fig:chivsT}(a) depicts the ensuing  open quantum dynamics. The impurity level shift triggers electron transfer from the reservoirs to the impurity on a relatively short timescale. Subsequently, reservoir electrons redistribute their spins to screen the impurity's localized spin, leading to the formation of Kondo states at the impurity-reservoir interfaces over a longer timescale \cite{anders2005real,PhysRevLett.119.156601,Din24174120}. 
We denote $t_{\rm sh}$ and $t_{\rm lo}$ as representative times for the short-time and long-time regions, respectively. 

In this case, the real part of $\gamma_j^{\sigma}$, which represents the inverse lifetime (or decay rate) of the corresponding dissipaton level, depends exclusively on the decay channel index $p$ contained within the composite index $j=\{\sigma \nu s p\}$. Consequently, we shall compactly label it as $\operatorname{Re}(\gamma_p)$ in the subsequent sections.


Figures~\ref{fig:chivsT}(b)-(e) illustrate the time evolution of $\chi_m(t)$ ($m=0,1,2,3$) at different temperatures: $k_{\rm B}T = 3.0\,\Gamma$, $2.0\,\Gamma$, $1.0\,\Gamma$, and $0.6\,\Gamma$, respectively. As the temperature decreases, the components $\chi_m$ ($m\geqslant1$), particularly $\chi_3$, exhibit a pronounced increase, indicating a substantially higher precision requirement in the low-temperature regime. This trend is further corroborated by Table~\ref{tab:chivsT}, which lists the specific values of $\chi_m(t)$ at the asymptotic time $t_{\rm asy}=20\,\Gamma^{-1}$, demonstrating the heightened sensitivity and precision demands at lower temperatures. For the various temperatures investigated in Fig.~\ref{fig:chivsT}, the corresponding numerical parameters, including the number of \Pade poles ($N_{\text{Padé}}$) employed for decomposing the hybridization correlation functions, the total number of poles ($N_{\rm M}$), the number of dissipaton levels ($N_{\rm E}$), and the slowest decay rate of dissipatons ($\operatorname{Re}(\gamma)_{\rm min}$), are summarized in Table~\ref{tab:poleTs}.




\subsection{Physical origins of $\chi$}
\label{subsec:phys}

As summarized in Table~\ref{tab:poleTs}, a decrease in temperature leads to an expansion of the dissipaton space (characterized by a larger $N_{\rm E}$) alongside a reduced decay rate of the dissipaton modes. In this section, we elucidate the primary physical cause that leads to the pronounced increase in $\chi_m$ ($m\geqslant 1$) by disentangling these two interconnected factors.

\begin{figure}[t]
\includegraphics[width=\columnwidth]{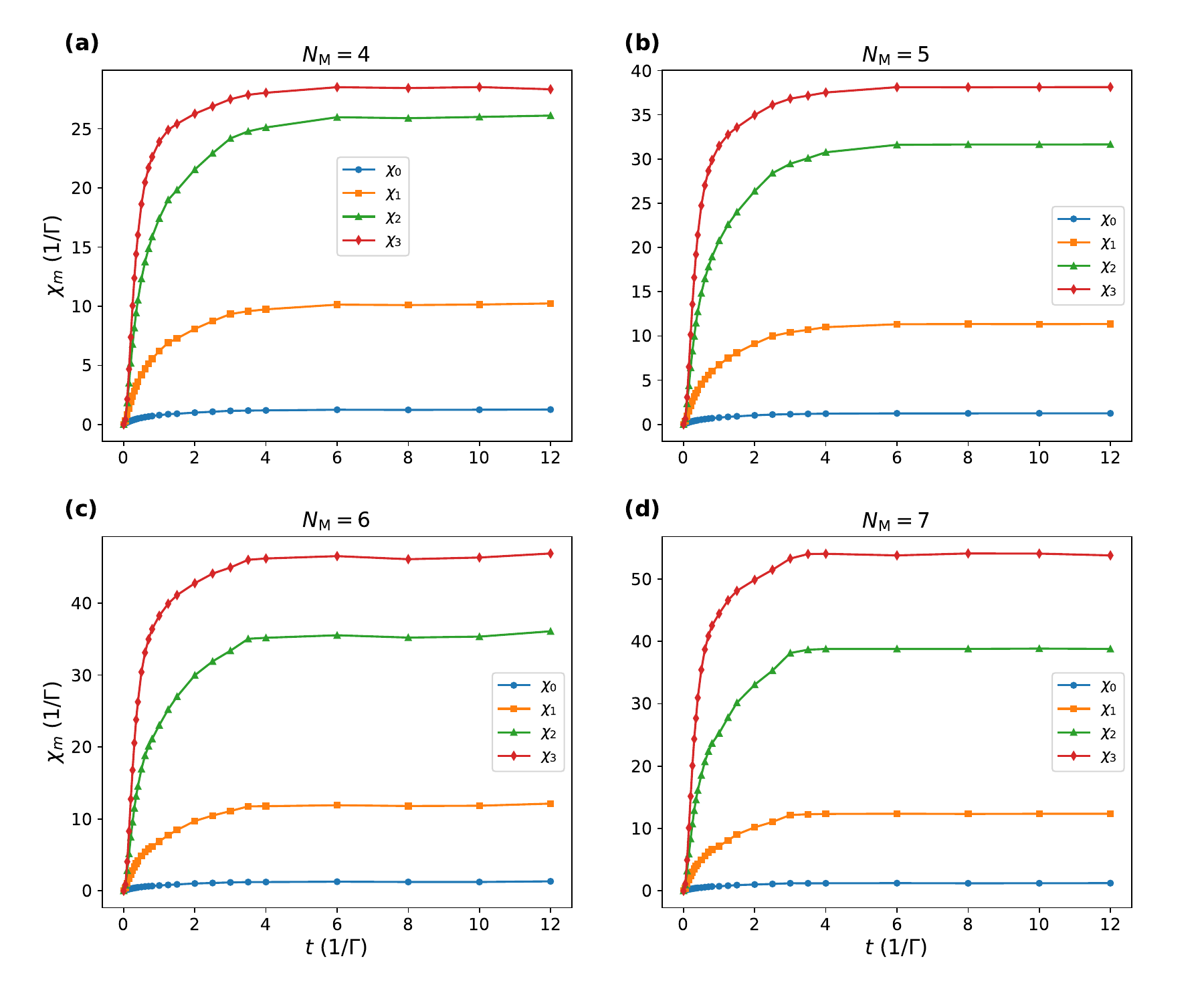}
\caption{Panels (a)–(d) depict  $\chi_m(t)\ (m=0,1,2,3)$ at $k_{\rm B}T = 3.0\, \Gamma$ for different numbers of decay channels,  $N_{\rm M}=4$, $5$, $6$, and $7$, respectively. 
System parameters (in units of $\Gamma$): $\epsilon_0 = U_0/2 = 2$, $\Delta\epsilon = -7$, and $\Delta U = 6$.
    }  \label{fig:chivsM}
\end{figure}


Firstly, we calculate the time-dependent $\chi_m(t)$ at $k_{\rm B}T = 3.0\,\Gamma$ for various numbers of decay channels $N_{\rm M}$, which correspond to different dimensions of the dissipaton space, $N_{\rm E}=8N_{\rm M}$, in this specific case. Figures~\ref{fig:chivsM}(a)-(d) illustrate the time evolution of $\chi_m(t)$ ($m=0,1,2,3$) under these configurations ($N_{\rm M}=4$, $5$, $6$, and $7$, respectively). As the dissipaton space expands, the sub-components $\chi_m$ ($m\geqslant 1$) exhibit only marginal increases by a mere factor of 2, which stands in sharp contrast to the hundred-fold increase observed in Fig.~\ref{fig:chivsT}. This clearly demonstrates that the expansion of the dissipaton space is not the predominant driver behind the pronounced growth of $\chi_m$ ($m\geqslant 1$) observed at lower temperatures. 
 
\begin{figure}[t]
\includegraphics[width=\columnwidth]{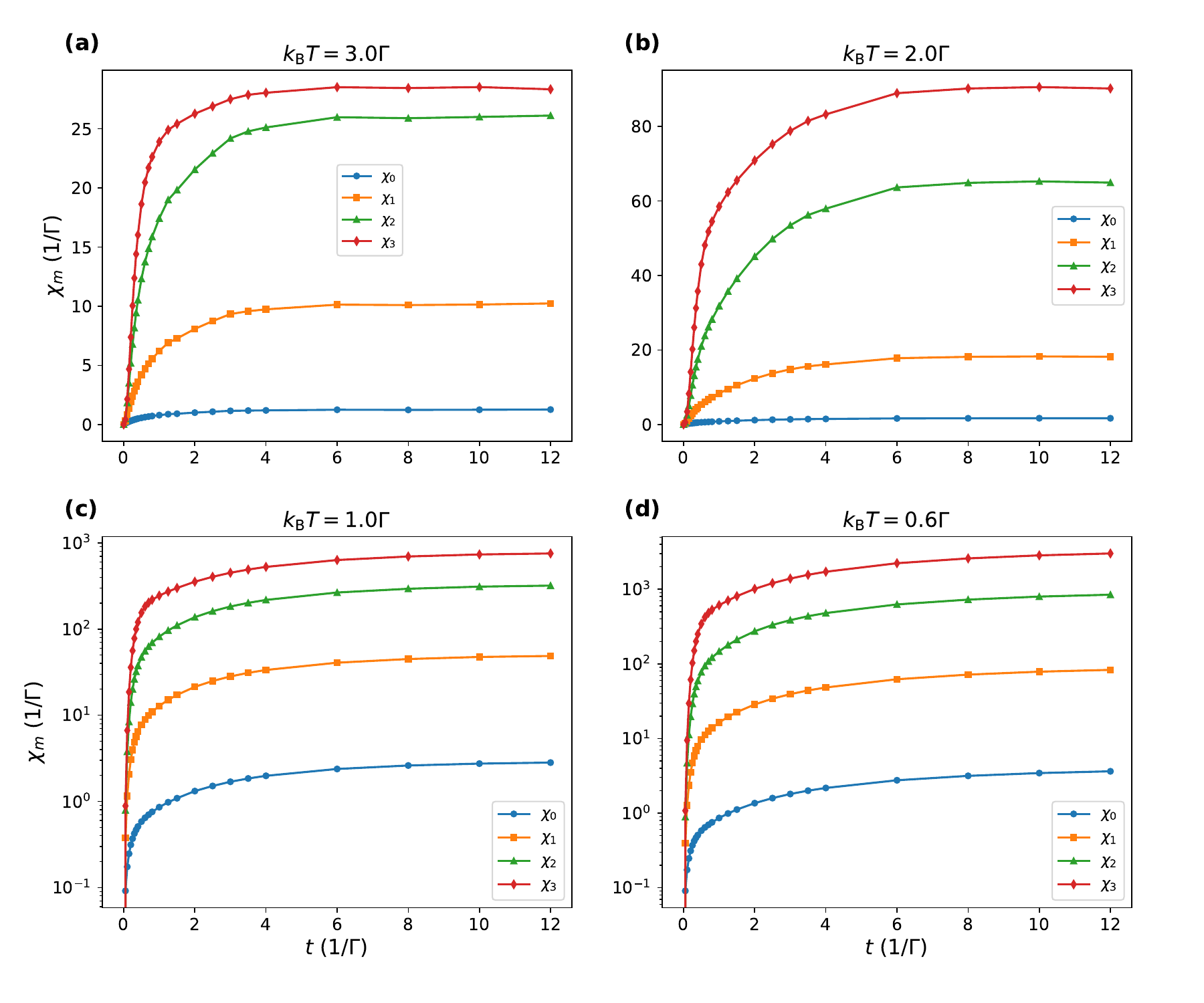}
\caption{
Panels (a)–(d) display $\chi_m(t)\ (m=0,1,2,3)$ at $k_{\rm B}T = 3.0\,\Gamma$, $2.0\,\Gamma$, $1.0\,\Gamma$, and $0.6\,\Gamma$, respectively, with $N_{\Pade}=4$ decomposition poles and a dissipaton space of size $N_{\rm E}=32$. The vertical axes in (c) and (d) use a logarithmic scale.
System parameters (in units of $\Gamma$): $\epsilon_0 = U_0/2 = 2$, $\Delta\epsilon = -7$, and $\Delta U = 6$.
}  \label{fig:chivsT_m4}
\end{figure}

\begin{table}[ht]
\centering
\caption{The decay rates of different dissipaton levels in Figure~\ref{fig:chivsT_m4}.}  \label{tab:gammam4}
\begin{tabular*}{\columnwidth}{@{\extracolsep\fill}lccccc}
\hline\hline
$k_{\rm B}T/\Gamma$  & $3.0$  & $2.0$  & $1.0$  & $0.6$   \\
\hline
$\operatorname{Re}(\gamma_0)/\Gamma$  & $9.42$  & $6.28$  & $3.14$  & $1.88$    \\
$\operatorname{Re}(\gamma_1)/\Gamma$  & $28.6$  & $19.1$  & $9.55$ & $5.73$  \\
$\operatorname{Re}(\gamma_2)/\Gamma$  & $50.0$ & $50.0$ & $27.2$ & $16.3$  \\
$\operatorname{Re}(\gamma_3)/\Gamma$ & $81.5$  & $54.4$  & $50.0$  & $50.0$ \\
\hline\hline
\end{tabular*}
\end{table}

Secondly, we calculate the time-dependent $\chi_m(t)$ at different temperatures while fixing the number of decay channels to $N_{\rm M} = 4$ (correspondingly, $N_{\text{Padé}} = 3$), which maintains a constant dimension of the dissipaton space ($N_{\rm E}=32$) across all cases. Figure~\ref{fig:chivsT_m4}(a)-(d) illustrate the time evolution of $\chi_m(t)$ ($m=0,1,2,3$) at $k_{\rm B}T = 3.0\,\Gamma$, $2.0\,\Gamma$, $1.0\,\Gamma$, and $0.6\,\Gamma$, respectively. As the temperature decreases, the sub-components $\chi_m$ ($m\geqslant 1$) still exhibit a pronounced increase, fully consistent with the hundred-fold growth observed in Fig.~\ref{fig:chivsT}. With a fixed $N_{\rm M}$, the lowering of temperature manifests directly as a proportional reduction in the decay rates of the dissipaton modes, as summarized in Table~\ref{tab:gammam4}. This unambiguously demonstrates that the reduction in the dissipaton decay rates, rather than the expansion of the state space, is the predominant driver behind the substantial growth of $\chi_m$ ($m\geqslant 1$) at low temperatures. 


\subsection{Channel decomposition of $\chi$}
While the Sec.~\ref{subsec:phys} demonstrates that the reduction in dissipaton decay rates drives the anomalous escalation of $\chi_m$ at lower temperatures, we microscopically substantiate this conclusion here by comparing the contributions of distinct decay channels.

To assess the contribution of different decay channels to the non-Markovianity, we distribute the non-Markovian susceptibility $\chi$ to different channels 
and define $\chi_p(t)$ as
\begin{equation}
    \chi_p(t)= \sum_{m_{j_p}=1}
\left\Vert\lla\vec{0}\vert\int_{0}^{t} dz \, e^{\mathcal{L}(t-z)} \vert \vec{m}\rra_{_{\rm E}}
    \right\Vert_{F} \cdot \frac{\sum_{j_p} m_{j_p}f_{j_p}}{\sum_j m_jf_j},
\end{equation}
where $j_p$ is the $j=\{\sigma \nu s p\}$ index that is related to the $p$-th channel, and $f_j$ is the weight of the level labeled by $j$. The factor $\frac{\sum_{j_p} m_{j_p}f_{j_p}}{\sum_j m_jf_j}$ thus quantifies the relative contribution of the $p$-th channel.

To determine $f_j$, we consider the $M_{\rm max}=1$ case, where the contributions from different dissipaton levels do not mix. And we can determine $f_j$ by
\begin{equation}
    f_j =f_p = \frac{\chi_{p,1}}{\sum_p \chi_{p,1}},
\end{equation}
where $j=\{\sigma \nu s p\}$ and 
$\chi_{p,1} = \chi_p(t=t_{\rm asy},M_{\rm max}=1)$ is the contribution of the $p$-th channel. Noting that
\begin{equation}
    \chi_p(t,M_{\rm max}=1) = \sum_{m_{j_p}=1}
    \left\Vert\lla\vec{0}\vert\int_{0}^{t} dz \, e^{\mathcal{L}(t-z)} \vert \vec{m}\rra_{_{\rm E}}\right\Vert_{F},
\end{equation}
we see that this quantity is completely independent of $f_j$.

Therefore, we have $\sum_m\chi_m = \sum_p \chi_p$, and the magnitude of $\chi_p$ reflects the contribution of $p$-th channel to the total error susceptibility $\chi$.

\begin{figure}[t]
\includegraphics[width=\columnwidth]{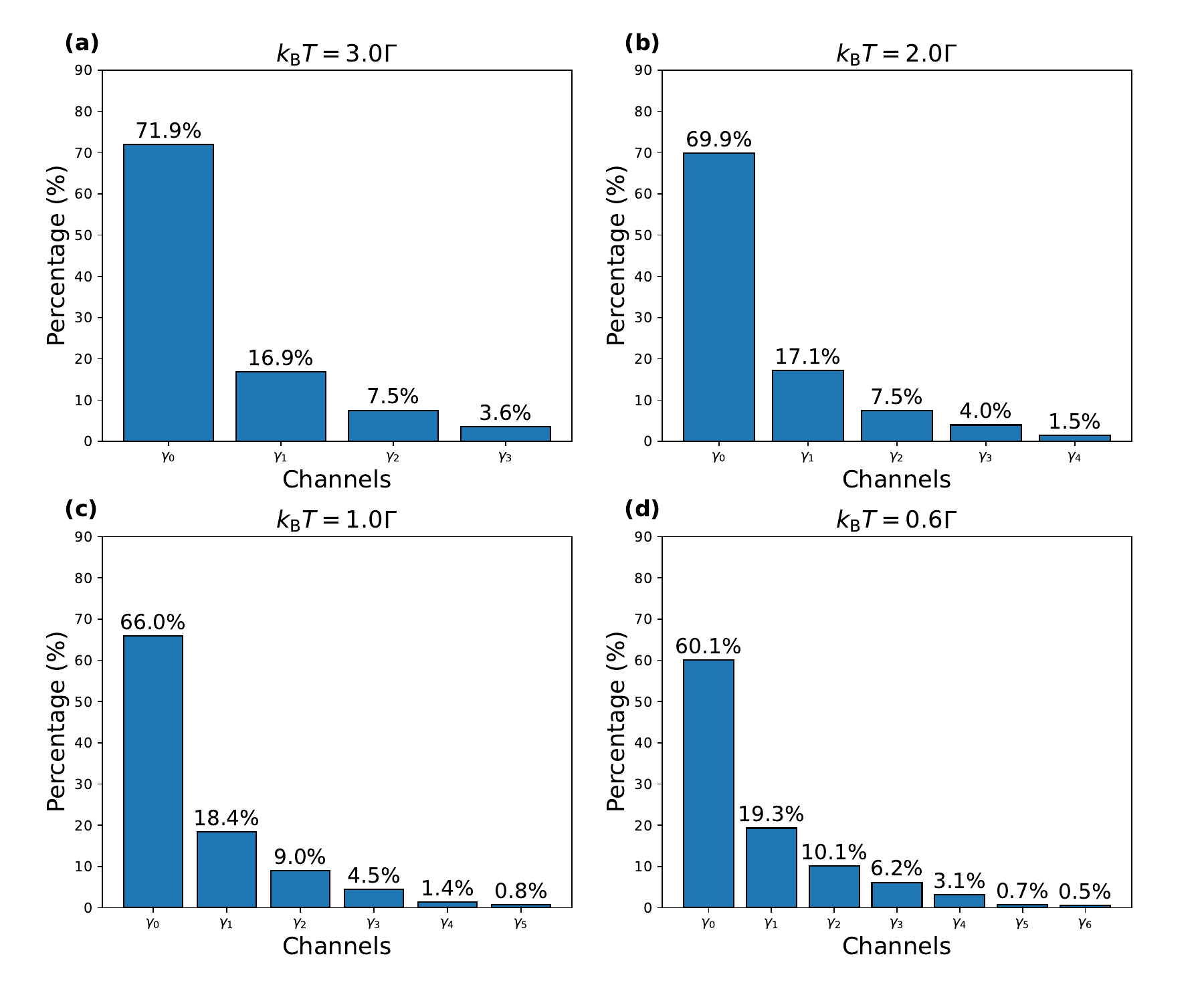}
\caption{Contribution to the susceptibitly $\chi$ from the $p$-th decay channel $\gamma_p$ at $t_{\rm asy} = 20\,\Gamma^{-1}$ for (a) $k_{\rm B}T = 3.0\,\Gamma$, (b) $2.0\,\Gamma$, (c) $1.0\,\Gamma$, and (d) $0.6\,\Gamma$, respectively, with the decay rates $\operatorname{Re}(\gamma_0) < \operatorname{Re}(\gamma_1)< \cdots <\operatorname{Re}(\gamma_{N_{_{\rm M}}-1})$.
The percentages are calculated from the channel contribution $\chi_p(t_{\rm asy})$, showing how the total susceptibility is distributed among individual dissipaton channels.
    }  \label{fig:chd}
\end{figure}

Figure~\ref{fig:chd} shows the contribution of different channels at $t_{\rm asy}=20\,\Gamma^{-1}$ for the various temperatures examined in Fig.~\ref{fig:chivsT}, where percentages are calculated from the channel contribution $\chi_p(t_{\rm asy})$ (see Appendix~\ref{app:tabs} for details).  The results explicitly demonstrate that the slowest-decaying mode provides the dominant contribution, thereby verifying that the mitigation of dissipaton decay rates is the primary driver behind the escalation of $\chi$.


This conclusion is highly non-trivial, as it unambiguously establishes that the precision degradation stems inherently from dynamical non-Markovian memory effects, rather than from a strong static system-environment coupling. To further contrast this mechanism, one might consider a scenario where a partial trace is performed over certain internal degrees of freedom within a closed system, which formally leads to an error analysis analogous to Sec.~\ref{subsec:error}. Crucially, however, our numerical findings confirm that the resulting susceptibility in such a closed sub-system should remain entirely trivial, precisely because it lacks genuine environmental dissipation channels.

\subsection{System size dependence of $\chi$}
Having elucidated the dependence of $\chi$ on environmental degrees of freedom, we proceed to investigate its scaling behavior with respect to the system size $\NS$. To adapt to affordable computational overhead, we employ a model of a localized impurity coupled to a single noninteracting electron reservoir at $k_{\rm B}T = 0.6\,\Gamma$, with a decay channel number of $N_{_{\rm M}}=3$. While this specific $N_{_{\rm M}}$ is insufficient to yield fully converged, numerically accurate results, it serves as an optimal truncation that qualitatively captures the essential non-Markovian features while minimizing the size of dissipaton configuration space.

The impurity Hamiltonian is 
\begin{align}
H_{_{\rm S}}(t) &= \sum_{i=1}^{\NS/2}H_{_{i}}(t) 
\notag \\
&=\sum_i \Big( \epsilon_{0}\hat{n}_{i\uparrow} + \epsilon_{0}\hat{n}_{i\downarrow}  + U_0\hat{n}_{i\uparrow}\hat{n}_{i\downarrow} +\Theta(t-t_0)
\notag  \\
& \qquad
  \times \big[\Delta \epsilon (\hat{n}_{i\uparrow} + \hat{n}_{i\downarrow}) + \Delta U\hat{n}_{i\uparrow}\hat{n}_{i\downarrow} \big] \Big).
\end{align}

\begin{figure}[t]
\includegraphics[width=\columnwidth]{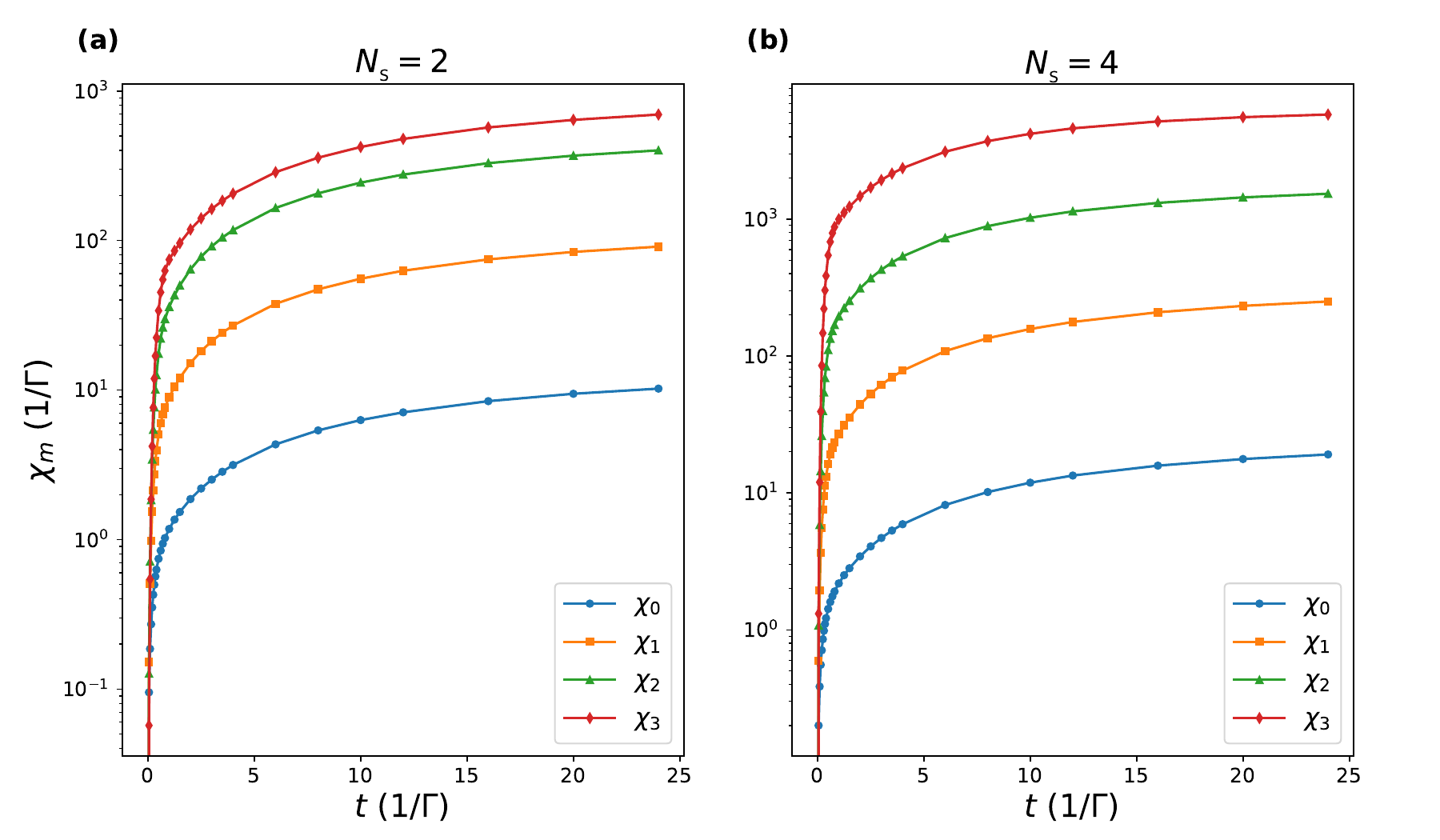}
\caption{$\chi_m(t)\ (m=0,1,2,3)$ at $k_{\rm B}T = 0.6\,\Gamma$ for (a) $\NS=2$ and (b) $\NS=4$, respectively, with $N_{\rm M}=3$ decay channels. 
System parameters (in units of $\Gamma$): $\epsilon_0 = U_0/2 = 2$, $\Delta\epsilon = -7$, and $\Delta U = 6$.
    }  \label{fig:chivsNs}
\end{figure}

\begin{table}[ht]
\centering
\caption{The value of $\chi_m(t)$ at $t_{\rm asy}=100\,\Gamma^{-1}$ for $k_{\rm B}T = 0.6\,\Gamma$ with different system size $\NS$.}  \label{tab:chivsNs}
\begin{tabular*}{\columnwidth}{@{\extracolsep\fill}lccccc}
\hline\hline
$\NS$  & $2$  & $4$  & $6\ (M_{\rm max=2})$   \\
\hline
$\chi_0(t_{\rm asy})/(1/\Gamma)$  & $1.26\times 10^{1}$  & $2.34\times 10^{1}$  & $4.33\times 10^{1}$  \\ 
$\chi_1(t_{\rm asy})/(1/\Gamma)$  & $1.12\times 10^{2}$  & $3.05\times 10^{2}$  & $7.19\times 10^{2}$  \\ 
$\chi_2(t_{\rm asy})/(1/\Gamma)$  & $4.95\times 10^{2}$  & $1.79\times 10^{3}$  & $5.71\times 10^{3}$   \\ 
$\chi_3(t_{\rm asy})/(1/\Gamma)$   & $8.61\times 10^{2}$  & $6.40\times 10^{3}$  & / \footnote{The total Liouville space size with $\NS=6$ and $M_{\rm max=3}$ exceeds the memory available on our current local machine.}   \\
\hline\hline
\end{tabular*}
\end{table}


Figures~\ref{fig:chivsNs}(a) and (b) illustrate the time evolution of $\chi_m(t)$ ($m=0,1,2,3$) at $k_{\rm B}T = 0.6\,\Gamma$ for $N_{\rm S}=2$ and $N_{\rm S}=4$, respectively. The corresponding limit values $\chi_m(t_{\rm asy})$ are summarized in Table~\ref{tab:chivsNs}. As the system size expands, the susceptibility $\chi_m$ ($m\geqslant 1$) exhibits a pronounced increase, demonstrating more stringent precision requirements in larger systems. This scaling behavior stems directly from the algebraic relation $N_{\rm E}=2N_{\rm M}N_{\rm S}$: an increase in $N_{\rm S}$ inevitably leads to a proportional growth in $N_{\rm E}$ by increasing the number of levels in each decay channel. Consequently, the proliferation of slow-decaying dissipaton levels drives the drastic escalation of the susceptibility $\chi$.


\section{Concluding remarks} \label{sec:conclude}


In this work, we establish a comprehensive theoretical framework, complemented by converged numerical validation, to systematically uncover the physical mechanisms underlying the precision bottlenecks in variational non-Markovian dynamics. By dissecting how the non-Markovian susceptibility $\chi$ structurally depends on both system degrees of freedom and environmental statistical properties, we elucidate the core error amplification pathway: non-local error backflow from the environment conveys not only physical information but also accumulated numerical errors. Specifically, our analysis identifies two key factors that amplify the error: the proliferation of slow-decaying dissipaton levels with increasing system size, and the reduction of dissipaton decay rates at lower temperatures, which together lead to a sharp increase in $\chi$. 
Crucially, our analysis demonstrates that this precision bottleneck is by no means an artifact confined to variational fermionic DQME; rather, it highlights a generic, intrinsic challenge across bosonic environments and broader numerical non-Markovian approaches.

This generic error-backflow mechanism gives rise to a formidable, systemic challenge for simulating complex many-body open quantum systems: a severe conflict between increasingly tight precision requirements for variational ansatzes (driven by rising $\chi$ values in strong-memory regimes) and the concomitant amplification of stochastic sampling errors in large-scale dynamics. To preserve physical fidelity, variational ansatzes must meet far stricter precision standards as non-Markovian memory grows. However, stochastic sampling fluctuations inherently scale unfavorably with increasing system size, causing the signal-to-noise ratio to degrade rapidly. This limitation renders brute-force scaling of variational methods computationally infeasible, restricting current NQS approaches to small systems and weakly non-Markovian regimes.

To resolve this fundamental conflict, future variational frameworks must move beyond brute-force scaling toward a dual-pronged strategy. On the architectural front, rescaling the dissipaton basis provides a promising route to inherently suppress the growth of $\chi$ in strong-memory regimes, thereby reducing the required precision threshold at the expense of necessitating more expressive variational representations. Alongside these advances, novel variance-reduced sampling protocols and physics-guided proposal distributions will be required to preserve high signal-to-noise ratios in large-scale simulations. Finally, a fine-grained theoretical framework is needed to disentangle distinct error sources, including approximation, regularization, and sampling errors, laying the groundwork for targeted mitigation strategies tailored to each specific error mechanism.

Ultimately, by diagnosing the exact failure mechanisms of current methods, this work provides a solid theoretical foundation and a clear roadmap for conquering the large-scale, strongly non-Markovian frontier.

\begin{acknowledgments}

Support from the Quantum Science and Technology -- National Science and Technology Major Project (Grant Nos.\ 2021ZD0303301 and 2021ZD0303306), the National Natural Science Foundation of China (Grant Nos.\ 22393912, 22425301, 22373091 and 22573099), and the AI for Science Foundation of Fudan University (Grant No. FudanX24AI023) is gratefully acknowledged. 
%


\end{acknowledgments}



\section*{Data availability}
The raw and processed data required to reproduce these findings 
and the source code for the numerical solver developed in this work are available at \url{https://github.com/caolong-cn/NQS-DQME_fermions}.

\appendix
\section{Time-dependent Liouvillian}
\label{app:tl}
We can easily generalize the theoretical analysis in Sec.~\ref{subsec:error} to time-dependent situations by two simple 
modifications.

Firstly, we replace the propagator $e^{\mathcal{L}(t-z)}$ in Eq.~\ref{eq:deltarho} with the time-ordered exponential $\mathcal{T}\left\{\exp\left[\int_z^t \mathcal{L}(\tau) d\tau\right]\right\}$.

Secondly, we replace the second Cauchy-Schwarz inequality $\Vert Ax\Vert_2 \leqslant \Vert A\Vert_F \Vert x\Vert_2$ with $\Vert Ax\Vert_2 \leqslant \Vert A\Vert_2 \Vert x\Vert_2$. Here, $\Vert A\Vert_2$ is the 2-norm (spectral norm) of the $\NS^2\times \NS^2$ matrix $A$, where $A$ is
\begin{equation}
A=\lla \vec{0} \vert \int_{0}^{t} dz \, \mathcal{T}\left\{\exp\left[\int_z^t \mathcal{L}(\tau) d\tau\right]\right\}\vert\vec{m}\rra_{_{\rm E}}.
\end{equation}
By definition, the induced matrix 2-norm (spectral norm) of a matrix $A$ is
\begin{equation}
\Vert A \Vert_2 = \sigma_{\max}(A),  
\end{equation}
where $\sigma_{\max}(A)$ is the largest singular value of $A$. And the equality condition for the Cauchy-Schwarz inequality $\Vert Ax\Vert_2 \leqslant \Vert A\Vert_2 \Vert x\Vert_2$ does not require $A$ to be a rank-1 matrix.

\section{Regularization}
\label{app:reg}
We calculate $\chi(t)$ by analytically integrating $\int_{0}^{t} dz \, e^{\mathcal{L}(t-z)}$, 
\begin{equation}
    \left\Vert\lla\vec{0}\vert\int_{0}^{t} dz \, e^{\mathcal{L}(t-z)} \vert \vec{m}\rra_{_{\rm E}}
    \right\Vert_{F}
    = 
     \left \Vert\lla\vec{0}\vert\mathcal{L}^{-1} 
     (e^{\mathcal{L}t}-1)\vert \vec{m}\rra_{_{\rm E}}
    \right\Vert_{F}.
\end{equation}

To calculates $\mathcal{L}^{-1} 
     (e^{\mathcal{L}t}-1)$
we have to regularize the $\mathcal{L}$ in the denominator. We have two regularization strategies which can induce different errors.
To estimate the errors from regularization, we expand this quantity as
\begin{equation}
    \mathcal{L}^{-1} 
     (e^{\mathcal{L}t}-1) = \sum_k \frac{e^{\lambda_k t} - 1}{\lambda_k} \vert \psi_k \rra \lla \phi_k \vert,
\end{equation}
where the $\vert \psi_k \rra$ and $\vert \phi_k \rra $ are the $k$-th right and left eigenvectors of $\mathcal{L}$, and $\lambda_k$ is the corresponding singular value.

Firstly, we replace the $L$ with $L-\epsilon I$ in both the denominator and the numerator, and the results are denoted by $[\chi_m]_1$. This regularization will induce errors:
\begin{enumerate}
    \item The steady state: replacing $\frac{e^{\lambda_0 t} - 1}{\lambda_0} =t$ by $\frac{e^{-\epsilon t} - 1}{-\epsilon} $. When $t\ll 1/\epsilon$, this error is small. When 
    $t> 1/\epsilon$, the new term tends to $1/\epsilon$ rather than scales linearly with $t$.
    
    \item Other states: replacing $\frac{e^{\lambda_k t} - 1}{\lambda_k} =t$ by $\frac{e^{(\lambda_k-\epsilon) t} - 1}{\lambda_k-\epsilon} $. When $\epsilon\ll\operatorname{Re} ( \lambda_k )$, this error is small.
\end{enumerate}

Secondly, we only replace the $L$ with $L-\epsilon I$ in the denominator, and the results are denoted by $[\chi_m]_2$. This regularization will induce errors:
\begin{enumerate}
    \item The steady state: replacing $\frac{e^{\lambda_0 t} - 1}{\lambda_0} =t$ by $\frac{1 - 1}{-\epsilon}=0 $. This regularization will remove the term linear to $t$.
    
    \item Other states: replacing $\frac{e^{\lambda_k t} - 1}{\lambda_k} =t$ by $\frac{e^{\lambda_k t} - 1}{\lambda_k-\epsilon} $. When $\epsilon\ll\operatorname{Re} ( \lambda_k )$, this error is small.
\end{enumerate}

Evidently, these two strategies differ exclusively in their steady-state behaviors, with their discrepancy characterized solely by a correction term linear in $t$,
\begin{align}
    \Delta \chi_m(t) &= [\chi_m(t)]_1-[\chi_m(t)]_2
    \notag \\
    &=t\cdot\sum_{\sum_j m_j =m}
    \left\Vert
    \lla\vec{0}
 \vert \psi_0 \rra_{_{\rm E}} \lla \phi_0 
    \vert \vec{m}
    \rra_{_{\rm E}}
    \right\Vert_{F}.
    \label{eq:deltachi}
\end{align}


\begin{figure}
    \centering
    \includegraphics[width=\columnwidth]{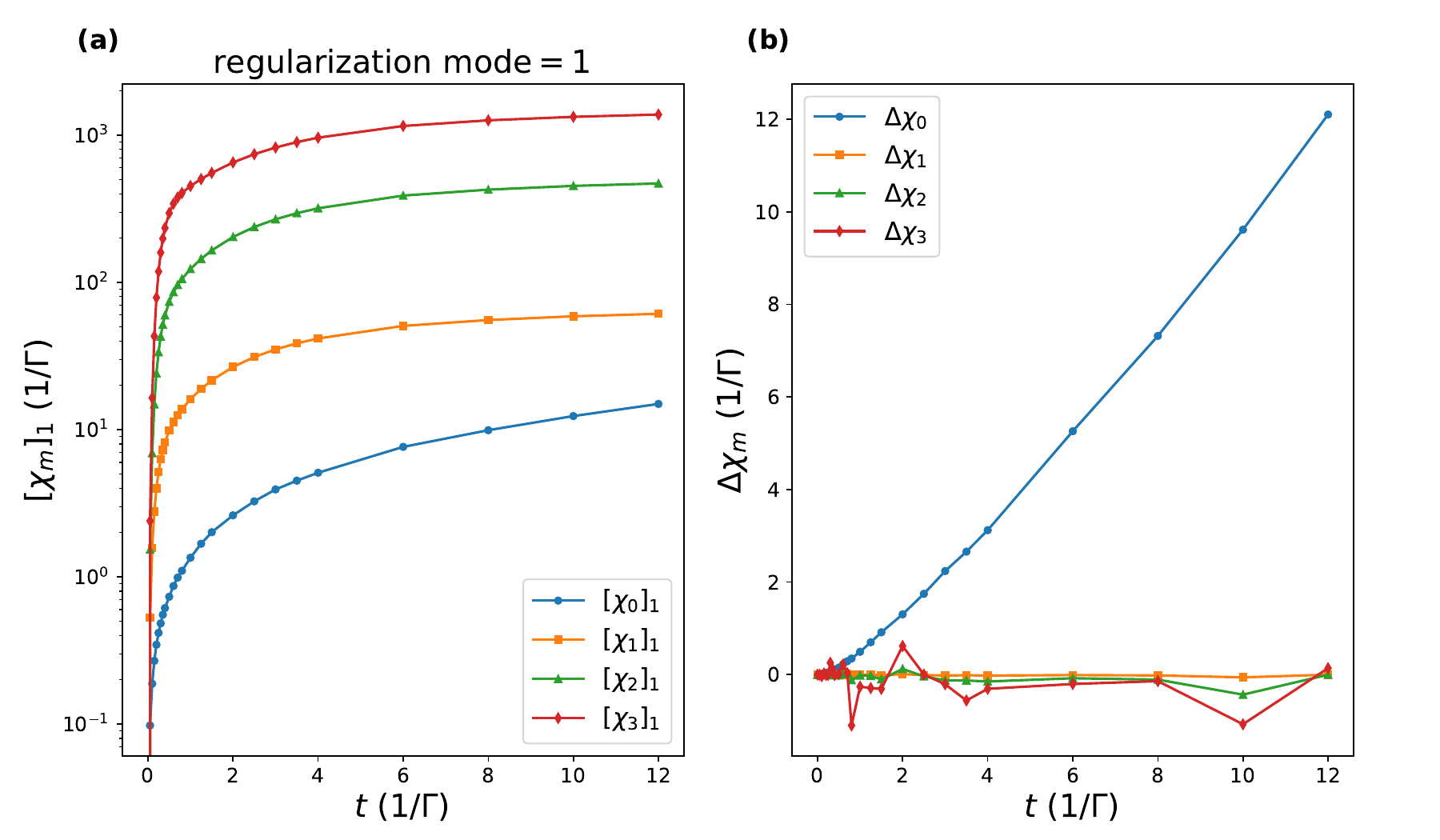}
    \caption{(a) $\chi_m(t)\ (m=0,1,2,3)$ at $k_{\rm B}T = 1.0\, \Gamma$ under the first regularization scheme. (b) $\Delta\chi_m(t)\ (m=0,1,2,3)$ at $k_{\rm B}T = 1.0\, \Gamma$, where $\Delta\chi_m(t)$ denotes the discrepancy between the two regularization schemes.
System parameters (in units of $\Gamma$): $\epsilon_0 = U_0/2 = 2$, $\Delta\epsilon = -7$, and $\Delta U = 6$.}
    \label{fig:chir}
\end{figure}

Figure~\ref{fig:chir}(a) displays the trajectories of $[\chi_m]_1\ (i=0,1,2,3)$ at $k_{\rm B}T = 1.0\, \Gamma$, while the corresponding discrepancies, defined as $\Delta\chi_m(t)=[\chi_m]_1-[\chi_m]_2$, are depicted in Fig.~\ref{fig:chir}(b). Crucially, Fig.~\ref{fig:chir}(b) demonstrates that the two regularizations differ in $\chi_0$ exclusively by a secular term linear in $t$, which is consistent with our prior theoretical derivation. For the components $\chi_m\ (m\geqslant 1)$, the two schemes yield nearly identical results, revealing that the expansion coefficients in Eq.~\eqref{eq:deltachi} vanish for $m\geqslant1$. This numerical confirmation further underscores that the matrix element $\lla \phi_0 \vert\vec{m}\rra_{_{\rm E}}$ remains negligibly small for $\vec{m}\neq\vec{0}$, thereby substantiating the physical validity of the second Cauchy-Schwarz inequality.

The second regularization scheme can exclude the trivial effect from the steady state. So when investigating the non-Markovian effect, we choose the second regularization scheme.


\section{Supplementary tables}
\label{app:tabs}
\begin{table}[ht]
\centering
\caption{The value of $\chi_m(t)$ at $t_{\rm asy}=20\,\Gamma^{-1}$ for $k_{\rm B}T = 3.0\,\Gamma$ with various decay channels $N_{\rm M}$.}  \label{tab:chivsM}
\begin{tabular*}{\columnwidth}{@{\extracolsep\fill}lccccc}
\hline\hline
$N_{\rm M}$   & $4$  & $5$  & $6$  & $7$\\
\hline
$\chi_0(t_{\rm asy})/(1/\Gamma)$    & $1.24$  & $1.25$  & $1.24$  & $1.22$ \\
$\chi_1(t_{\rm asy})/(1/\Gamma)$    & $1.01\times 10^{1}$  & $1.13\times 10^{1}$  & $1.20\times 10^{1}$  & $1.24\times 10^{1}$ \\
$\chi_2(t_{\rm asy})/(1/\Gamma)$    & $2.60\times 10^{1}$  & $3.16\times 10^{1}$  & $3.59\times 10^{1}$  & $3.88\times 10^{1}$ \\
$\chi_3(t_{\rm asy})/(1/\Gamma)$     & $2.85\times 10^{1}$  & $3.82\times 10^{1}$  & $4.69\times 10^{1}$   & $5.41\times 10^{1}$\\
\hline\hline
\end{tabular*}
\end{table}

\begin{table}[ht]
\centering
\caption{Number of \Pade poles and environmental degrees of freedom in Figure~\ref{fig:chivsM}.}  \label{tab:poleMs}
\begin{tabular*}{\columnwidth}{@{\extracolsep\fill}lccccc}
\hline\hline
$N_{\rm M}$  & 4  & 5  & 6 & 7  \\
\hline
$N_{\Pade}$  & 3  & 4  & 5  & 6    \\
$N_{\rm E}$  & 32 & 40 & 48 & 56  \\
$\operatorname{Re}(\gamma)_{\rm min}/\Gamma$ & $9.42$  & $9.42$  & $9.42$  & $9.42$ \\
\hline\hline
\end{tabular*}
\end{table}

Table~\ref{tab:chivsM} lists the specific values of $\chi_m(t)$ at the asymptotic time $t_{\rm asy}=20\,\Gamma^{-1}$ in Figure~\ref{fig:chivsM}. For the various environmental configurations investigated in Figure~\ref{fig:chivsM}, the key numerical parameters are summarized in Table~\ref{tab:poleMs}.

\begin{table}[ht]
\centering
\caption{The value of $\chi_m(t)$ at $t_{\rm asy}=20\,\Gamma^{-1}$ for different temperatures with the same $N_{\rm E}=32$.}  \label{tab:chivsT_m4}
\begin{tabular*}{\columnwidth}{@{\extracolsep\fill}lccccc}
\hline\hline
$k_{\rm B}T/\Gamma$  & $3.0$  & $2.0$  & $1.0$  & $0.6$\\
\hline
$\chi_0(t_{\rm asy})/(1/\Gamma)$  & $1.24$  & $1.68$  & $2.93$  & $3.98$  \\
$\chi_1(t_{\rm asy})/(1/\Gamma)$  & $1.01\times 10^{1}$  & $1.83\times 10^{1}$  & $5.09\times 10^{1}$  & $9.13\times 10^{1}$ \\
$\chi_2(t_{\rm asy})/(1/\Gamma)$  & $2.60\times 10^{1}$  & $6.53\times 10^{1}$  & $3.33\times 10^{2}$  & $9.31\times 10^{2}$ \\
$\chi_3(t_{\rm asy})/(1/\Gamma)$   & $2.85\times 10^{1}$  & $9.07\times 10^{1}$  & $7.86\times 10^{2}$  & $3.33\times 10^{3}$ \\
\hline\hline
\end{tabular*}
\end{table}


\begin{table}[ht]
\centering
\caption{Number of \Pade poles and environmental degrees of freedom in Figure~\ref{fig:chivsT_m4}.}  \label{tab:poleTsm4}
\begin{tabular*}{\columnwidth}{@{\extracolsep\fill}lccccc}
\hline\hline
$k_{\rm B}T/\Gamma$  & $3.0$  & $2.0$  & $1.0$  & $0.6$   \\
\hline
$N_{\Pade}$  & 3  & 3  & 3  & 3    \\
$N_{\rm M}$  & 4  & 4  & 4 & 4  \\
$N_{\rm E}$  & 32 & 32 & 32 & 32  \\
$\operatorname{Re}(\gamma)_{\rm min}/\Gamma$ & $9.42$  & $6.28$  & $3.14$ & $1.88$  \\
\hline\hline
\end{tabular*}
\end{table}

Table~\ref{tab:chivsT_m4} presents the specific values of $\chi_m(t)$ at the asymptotic limit $t_{\rm asy}=20\,\Gamma^{-1}$ in Figure~\ref{fig:chivsT_m4}. For the various temperatures investigated in Figure~\ref{fig:chivsT_m4}, the corresponding numerical and dynamic parameters are compiled in Table~\ref{tab:poleTsm4}.

\begin{table}[htbp]
    \centering
    \caption{
    Channel-dependent susceptibility $\chi_p(t_{\rm asy})$ evaluated at asymptotic time $t_{\rm asy} = 20\,\Gamma^{-1}$ for various temperatures.
    }
    \label{tab:mode_contribution}
    \begin{tabular}{cccccc}
        \hline\hline
                 $k_{\rm B}T / \Gamma$ &  $N_{\rm M}$ & \, $p$\, &  $\operatorname{Re}(\gamma_p) / \Gamma$ &  $\chi_p(t_{\rm asy})/(1/\Gamma)$ 
        \\
        \hline
        \multirow{4}{*}{3.0} & \multirow{4}{*}{4} 
          & 0 & 9.42 & $4.65\times 10^{1}$ \\
          & & 1 & 28.6 & $1.09\times 10^{1}$ \\
          & & 2 & 50.0 & $4.87$  \\
          & & 3 & 81.5 & $2.34$  \\
        \hline
        \multirow{5}{*}{2.0} & \multirow{5}{*}{5} 
          & 0 & 6.28 & $1.56\times 10^{2}$ \\
          & & 1 & 18.8 & $3.81\times 10^{1}$ \\
          & & 2 & 33.2 & $1.66*10{1}$ \\
          & & 3 & 50.0 & $8.85$ \\
          & & 4 & 92.6 & $3.33$ \\
        \hline
        \multirow{6}{*}{1.0} & \multirow{6}{*}{6} 
          & 0 & 3.14 & $1.29\times 10^{3}$ \\
          & & 1 & 9.42 & $3.60\times 10^{2}$ \\
          & & 2 & 15.8 & $1.76\times 10^{2}$ \\
          & & 3 & 24.9 & $8.79\times 10^{1}$ \\
          & & 4 & 50.0 & $2.77\times 10^{1}$ \\
          & & 5 & 70.5 & $1.53\times 10^{1}$ \\
        \hline
        \multirow{7}{*}{0.6} & \multirow{7}{*}{7} 
          & 0 & 1.88 & $5.75\times 10^{3}$ \\
          & & 1 & 5.66 & $1.85\times 10^{3}$ \\
          & & 2 & 9.43 & $9.63\times 10^{2}$ \\
          & & 3 & 13.4 & $5.88\times 10^{2}$ \\
          & & 4 & 20.8 & $3.00\times 10^{2}$ \\
          & & 5 & 50.0 & $6.70\times 10^{1}$ \\
          & & 6 & 59.9 & $4.84\times 10^{1}$ \\
        \hline\hline
    \end{tabular}
\end{table}

Table~\ref{tab:mode_contribution} summarizes the quantitative values of $\chi_p(t_{\rm asy})$ at various temperatures corresponding to Fig.~\ref{fig:chivsT}.

\clearpage


\end{document}